\documentclass[twocolumn,aps,floats,showpacs,superscriptaddress]{revtex4-2}
\usepackage[latin3]{inputenc}
\usepackage[makeroom]{cancel}
\usepackage{graphicx}
\usepackage{amsmath}
\usepackage{amsfonts}
\usepackage{amssymb}
\usepackage{color}
\usepackage{graphicx}
\usepackage{appendix}
\usepackage{ulem}
\usepackage{bm}
\usepackage{dsfont}

\usepackage[dvipsnames]{xcolor}

\usepackage{ulem}

\begin{document}

\title{Nonlinear nanoelectromechanics of a movable Cooper-pair box}
\author{S. Park}
\altaffiliation{These authors contributed equally to this work}
\affiliation{Center for Trapped Ion Quantum Science, Institute for Basic Science (IBS), Daejeon 34126, Republic of Korea}
\author{A. Patra}
\altaffiliation{These authors contributed equally to this work}
\affiliation{Department of Physics, Indian Institute of Technology Kharagpur, Kharagpur 721302, West Bengal, India} 
\author{L. Y. Gorelik}
\affiliation{Department of Physics, Chalmers University of Technology, SE-412 96 G{\" o}teborg, Sweden}
\affiliation{Department of Physics, Hanyang University, Seoul 04763, Republic of Korea}
\affiliation{Research Institute for Natural Science and High Pressure, Hanyang University, Seoul, 04763, Republic of Korea }
\author{M. J. Park}
\affiliation{Department of Physics, Hanyang University, Seoul 04763, Republic of Korea}
\affiliation{Research Institute for Natural Science and High Pressure, Hanyang University, Seoul, 04763, Republic of Korea }
\author{H. C. Park}
\affiliation{Department of Physics, Pukyong National University, Busan 48513, Republic of Korea }
\author{R. I. Shekhter}
\affiliation{Department of Physics, Hanyang University, Seoul 04763, Republic of Korea}
\affiliation{Research Institute for Natural Science and High Pressure, Hanyang University, Seoul, 04763, Republic of Korea }
\affiliation{Department of Physics, University of Gothenburg, SE-412 96 G{\" o}teborg, Sweden}

\date{\today}

\begin{abstract}
We theoretically study the dynamics of a movable Cooper-pair box coupled to a normal-metal pillar using a semiclassical approach. We analyze the dynamical stability induced by the nonlinear nanoelectromechanical coupling between the mechanical motion and an inelastic Andreev tunneling through linear stability and bifurcation analyses. 
As a function of $\eta$, defined as the ratio of electrostatic energy to Josephson coupling energy, the system exhibits reentrant stability. 
At small $\eta$, the fixed point loses stability through a supercritical Hopf bifurcation, giving rise to self-sustained vibrations. 
With a further increase of $\eta$, a second critical point appears, at which the fixed point regains stability. We show that this second transition corresponds to an inverse subcritical Hopf bifurcation in the adiabatic regime and to an inverse supercritical Hopf bifurcation in the nonadiabatic regime. These results extend previous studies of adiabatic self-vibrations to the nonadiabatic regime and reveal a rich nonlinear dynamical phase diagram arising from the interplay between electronic and mechanical degrees of freedom in superconducting devices.
\end{abstract}
\maketitle

\section{Introduction}
Hybrid systems combining superconducting qubits with nanomechanical elements have been explored as platforms for quantum information processing and transduction between electrical and mechanical degrees of freedom~\cite{Cleland03,Roukes09,Sillanpaa13}. In most realizations, the coupling is mediated by phonon exchange and acts as a transducer between electrical excitations and mechanical motion. On the other hand, when the mechanical element is an active part of the device, as in a movable Cooper-pair box (CPB) coupled to a normal metal, electronic transport and mechanical motion are no longer separable but instead influence each other through a dynamical feedback, establishing a regime of nanoelectromechanics~\cite{Gorelik01,Gorelik98,Parafilo20,Fedorets05}. 

The movable CPB is a superconducting island whose mechanical motion modulates its Josephson coupling to a bulk superconductor~\cite{Park25}. This position-dependent coupling induces a Josephson force that depends on the CPB state and is distinct from an electrostatic force. When the island is additionally coupled to a normal metal, Andreev tunneling produces a current by converting pairs of electrons into Cooper pairs on the island, thereby changing its state. 
In particular, the coupling to a normal metal via inelastic Andreev tunneling provides pumping and dissipation for the CPB, leading to nonequilibrium dynamics and nonlinearities in its nanoelectromechanical behavior.   

Recently, it has been proposed that the CPB can exhibit self-sustained vibrations at low frequencies without external feedback~\cite{Park26}. 
In this setup, the CPB is attached to the free end of a normal-metal pillar and carries a current through inelastic Andreev tunneling in the presence of an electric field perpendicular to the current. The adiabatic vibrations arise from the noncommutativity of the Coulomb and Josephson couplings, with the instability supplied by inelastic Andreev tunneling. However, the phase diagram of the dynamics, including regimes beyond the adiabatic limit, remains largely unexplored. 

In this work, we study the phase diagram of the movable CPB introduced in Ref.~\cite{Park26} using linear stability and bifurcation analyses, and predict Hopf bifurcation points as a function of a parameter $\eta$, defined as the ratio of the electrostatic energy to the Josephson energy. The eigenvalue spectrum of the Jacobian matrix characterizes a sequence of dynamical transitions~\cite{Crawford}. At small $\eta$, the fixed point loses stability through a supercritical Hopf bifurcation, resulting in self-sustained vibrations whose vibrational amplitude increases as $\eta$ increases. As $\eta$ increases further, the system exhibits reentrant stability and reaches a second critical point at which the fixed point regains stability. We show that the character of the second transition depends on the adiabaticity of the CPB dynamics, determined by the ratio between the Andreev tunneling rate and the mechanical frequency. In the adiabatic regime, where the tunneling rate is much larger than the mechanical frequency, the transition corresponds to an inverse subcritical Hopf bifurcation, whereas in the nonadiabatic regime, it becomes an inverse supercritical Hopf bifurcation.

The paper is organized as follows. Section~\ref{model} is devoted to the formulation of the model for the movable CPB tunnel coupled to the normal-metal pillar and derive the equations for the reduced density matrix describing the dynamics of the CPB. Section~\ref{analysis} presents the Jacobian matrix using the linear stability analysis to classify the phase diagram with Hopf bifurcation critical points. Numerical results of the developed limit cycles in the adiabatic and nonadiabatic regimes are provided in Sec.~\ref{numerics}. A discussion is given in Sec.~\ref{discussion}.  

\section{Model Hamiltonian}\label{model} 
The Hamiltonian of our setup is 
\begin{equation}
\hat{H} = \hat{H}_{\text{CPB}} +\hat{H}_{\text{n}}+\hat{H}_{\text{A}} +\hat{H}_{\text{m}}. \label{htotal}
\end{equation}
The term $\hat{H}_{\text{CPB}}$ describes the CPB consisting of the superconducting island located in the vicinity of the origin in the $xy$ plane and coupled to the bulk superconductor via the Josephson energy $E_J(\hat{x})$, in the regime where the Coulomb blockade is removed by the gate voltage $V_G=e/C$ with $C$ the island capacitance, while the parity effect remains valid~\cite{Matveev93},   
\begin{equation}
\hat{H}_{\text{CPB}}= -E_{J}(\hat{x}) \hat{\sigma}_{1}+V(\hat{y}) \hat{\sigma}_3, \label{hcpb}
\end{equation}
where $\vec{\sigma}=(\hat{\sigma}_{1},\hat{\sigma}_{2},\hat{\sigma}_{3})$ denotes the vector of Pauli matrices acting on the qubit subspace spanned by the ground state $|0\rangle=(0,1)^T$ and the charged state with a single Cooper pair $|1\rangle = (1,0)^{T}$. 
$E_{J}(\hat{x})=E_J \exp(-\hat{x}/\lambda)$ represents the displacement-dependent Josephson coupling characterized by the tunneling length $\lambda$ between the CPB and the bulk superconductor along $x$ direction. The two side gates apply $V(\hat{y})=V_G+e\mathcal{E}\hat{y}$ in the $y$ direction, where $V_G$ shifts the electrostatic potential of the CPB and $\mathcal{E}$ is the electric field.  
The island is attached to the normal-metal pillar with the Hamiltonian,
\begin{equation}
\hat{H}_{\text{n}}=\sum_{k,\tau}(\varepsilon_k-e V_b) a_{k \tau}^{\dagger} a_{k \tau}, \label{hn} 
\end{equation}
where $a^{\dagger}_{k\tau}$ creates an electron of energy $\varepsilon_k$ with momentum $k$ and spin $\tau=\uparrow, \downarrow$ in the pillar. $V_b$ is the bias voltage. 
At the junction between the island and the pillar, electron exchange occurs via inelastic Andreev tunneling, which is modeled by~\cite{Park25}   
\begin{equation}
\hat{H}_{\text{A}}=t_{\text{A}}\sum_{k,k'}\left(a^{\dagger}_{k\uparrow}a^{\dagger}_{k'\downarrow}\hat{\sigma}^{-}+a_{k'\downarrow}a_{k\uparrow}\hat{\sigma}^{+}\right),  \label{ha} 
\end{equation}
where $\hat{\sigma}^{\pm}=(\hat{\sigma}_{1}\pm i\hat{\sigma}_{2})/2$.  This term takes into account the pair-electron process only under the assumption that the thermal broadening of the Fermi-Dirac distribution at temperature $T$ in the pillar is small compared to the superconducting gap, $k_B T \ll \Delta$, to suppress single-electron tunneling. We assume a constant tunneling coefficient $t_{\text{A}}$ in Eq.~\eqref{ha} during the mechanical motion. The mechanical degree of freedom of the combined CPB and the pillar, with frequency $\omega_0$, is described by  
\begin{equation}
\hat{H}_{\text{m}} = \frac{\hat{{\bf p}}^2}{2 m^*} +  \kappa\, \frac{\hat{{\bf r}}^2}{2}, \label{hm}
\end{equation}
where $\hat{{\bf r}}=\hat{x} {\bf{e}}_{x}+\hat{y} {\bf{e}}_y$ and ${\bf p}=\hat{p}_x {\bf{e}}_x +\hat{p}_y{\bf{e}}_y$ are position and momentum operators, respectively, and $\kappa=m^* \omega^2_0$ is the mechanical rigidity with the effective mass $m^{*}$.

We use a reduced density matrix approach to derive the dynamics of the CPB from the Liouville-von Neumann equation~\cite{Gorelik05} 
\begin{equation}
i \hbar \dot{\hat{\varrho}}_{\text{tot}}= [\hat{H},\hat{\varrho}_{\text{tot}}] \label{eom_1}
\end{equation}
for the total density matrix $\hat{\varrho}_{\text{tot}}$, where the overdot denotes time derivative.
To this end, we employ the Born-Markov approximation~\cite{Breuer}, which is applicable in the parameter regime, 
\begin{equation}
  \Delta > e V_b \gg k_B T,\, \hbar\Gamma,\, E_J, 
\end{equation}
where $\Gamma\equiv 2\pi t^2_{\text{A}} \nu^2|eV_b|/\hbar$, with $\nu$ the density of states in the pillar, is the tunneling rate between the CPB and the pillar. In this regime, any correlations within the pillar are quickly suppressed. Moreover, we employ a semiclassical approximation, treating the displacement as classical variables in the regime where the displacement scale is much larger than the zero-point fluctuation amplitude of the mechanical vibration, $\lambda \gg x_0=\sqrt{\hbar/m^*\omega_0}$. The coupling between the electronic and mechanical degrees of freedom can be characterized by  
\begin{equation}
  \epsilon \equiv  \frac{E_J}{ \kappa \lambda^2} = \left( \frac{x_0}{\lambda} \right)^2 \frac{E_J}{\hbar \omega_0}.  \label{em_coupling}
\end{equation}
This expression shows that the smallness of the zero-point fluctuation amplitude suppresses the coupling, provided that $E_J/(\hbar \omega_0)$ is not excessively large. Under the conditions, quantum correlations between the electronic and mechanical degrees of freedom remain negligible, which justifies the semiclassical treatment with the factorization of the density matrix as $\hat{\varrho}_{\text{tot}} = \hat{\varrho}_{\text{s}} \otimes\hat{\varrho}_{\text{m}} \otimes \hat{\varrho}^{\text{eq}}_{\text{n}}$, where $\hat{\varrho}_{\text{s}}$ and $\hat{\varrho}_{\text{m}}$ are the density matrices for the electronic states and mechanical motion of the CPB, respectively, and $\hat{\varrho}^{\text{eq}}_{\text{n}}$ is the equilibrium density matrix of the pillar. In the following, we consider the parameter range $E_J/(\hbar \omega_0) \sim 1-10^2$, covering both the nonadiabatic and adiabatic regimes. 

For convenience, we introduce dimensionless variables throughout the following analysis. Length, energy, and time are measured in units of $\lambda, \hbar \omega_0$, and $1/\omega_0$, respectively. Accordingly, the displacement is rescaled as $\hat{{\bf r}}/\lambda \rightarrow \hat{{\bf r}}$, and all energies and times below are understood in dimensionless units unless otherwise stated. Within the factorized description, the matrix $\hat{\varrho}_{\text{s}}$ is parameterized in the Bloch-sphere representation as 
\begin{equation}
  \hat{\varrho}_{\text{s}} = \frac{1}{2}\left(\hat{I}+ \bm{m}\cdot \vec{\sigma}\right),   
\end{equation}
where $\bm{m} = m_1 {\bf{e}}_1+m_2 {\bf{e}}_2+m_3 {\bf{e}}_3$ with $m_i=\rm{Tr}(\hat{\varrho}_{\text{s}} \hat{\sigma}_{i})$ denotes the corresponding Bloch vector.
Then Eq.~\eqref{eom_1} yields the equations for the displacement ${\bf r} = (1/\lambda)\text{Tr}(\hat{{\bf r}} \hat{\varrho}_{\text{m}})$ and the vector $\bm{m}$ in the following forms 
\begin{subequations}
    \begin{gather}
 \ddot{{\bf r}}+Q^{-1}\dot{{\bf r}}+ {\bf r}={\bf f}({\bf r}), \,\,\,\,\,\,\,\,  
{\bf f}({\bf r})= -\epsilon\,    \text{Tr}(\hat{\bf{F}}\hat{\varrho}_{\text{s}}), \label{eom2} \\
 \dot{\bm{m}}=\hat{B}\, \bm{m} + 2 \Gamma {\bf{e}}_3, \label{eom3}\\
 \hat{B} = 
 \begin{pmatrix}
   -\Gamma & -2 E_J \eta y & 0 \\
   2 E_J \eta y &  -\Gamma & 2 E_J e^{-x} \\
   0 &  -2 E_J e^{-x} & -2 \Gamma
 \end{pmatrix}, \nonumber
\end{gather}
\end{subequations}
where $\hat{\bf{F}}= e^{-x }\hat{\sigma}_1  {\bf e}_x + \eta\,\hat{\sigma}_3{\bf e}_y$ and $\eta=e\mathcal{E} \lambda/E_{J}$. $Q$ is the quality factor.

\section{Linear Stability and Bifurcation Analysis}\label{analysis}

In this section, we present the framework used to analyze the stability and bifurcation structure of the fixed points of the coupled CPB-oscillator semiclassical equations of motion. In dimensionless units, they read
\begin{subequations}
\begin{align}
    \dot{x} &= v_x, \\
    \dot{v}_{x} &= -x - \gamma v_{x} - \epsilon e^{-x} m_1, \\
    \dot{y} &= v_{y}, \\
    \dot{v}_y &= -y - \gamma v_{y} - \epsilon \eta m_3, \\
    \dot{\bm{m}} &= \hat{B}\, \bm{m} + 2 \Gamma {\bf{e}}_3,
\end{align}
\label{DL_EOM}
\end{subequations}
where $\gamma=Q^{-1}$ and the matrix $\hat{B}$ was introduced below Eq.~\eqref{eom3}. We denote the fixed point solution as a time-independent vector: $\bm{X}_{0} \equiv \left( x_0, v_{x0}, y_0, v_{y0}, \bm{m}_0 \right)$.

To analyze the stability of the fixed point $\bm{X}_0$ \cite{Crawford, Kuznetsov, Hilborn, Patra_1}, we linearize Eq.~\eqref{DL_EOM} about $\bm{X}_0$, obtaining the $7 \times 7$ Jacobian matrix $\mathbb{J}(\bm{X}_0)$. Its eigenvalues, referred to as the characteristic values, may be real or occur as complex-conjugate pairs. The corresponding eigenvectors define the characteristic directions. The fixed point is linearly stable when all characteristic values have negative real parts. Stability is lost when the real part of at least one characteristic value becomes positive. At the bifurcation point, the real part of one or more characteristic values crosses zero. The invariant subspaces spanned by the characteristic directions associated with characteristic values having negative, positive, and zero real parts are known as the stable, unstable, and center subspaces, respectively.

\begin{figure*}
    \centering
    \includegraphics[trim={0.75cm 0.55cm 0cm 0cm},clip,scale=0.85]{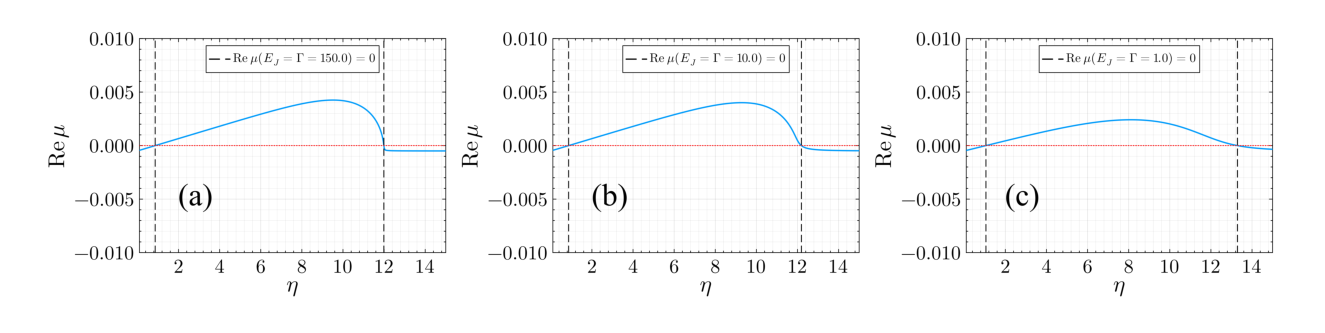}
    \caption{ Real parts of the critical eigenvalues of the fixed point $\bm{X}_0$ (solid blue) as a function of $\eta$. The red dotted line indicates $y = 0$ and is included as a guide to the eye. Panels (a) -- (c) correspond to $E_{J} (= \Gamma) = 150.0, 10.0,$ and $1.0$, respectively. We identify two Hopf bifurcations at $\eta_{c1}$ and $\eta_{c2}$, with $\eta_{c1} \lesssim 2.0$ and $\eta_{c2} \gtrsim 11.0$. The fixed point loses stability at $\eta_{c1}$ and regains stability at $\eta_{c2}$, resulting in a reentrant stability region. }
    \label{Fig:CPB_Jac}
\end{figure*}

All bifurcations reported in this work exhibit the same spectral signature. Near criticality, the Jacobian $\mathbb{J}(\bm{X}_0)$ possesses three complex-conjugate eigenvalue pairs and one real eigenvalue. Prior to the bifurcation, all eigenvalues have negative real parts. At criticality, a single complex-conjugate pair crosses the imaginary axis and acquires a positive real part, while the remaining eigenvalues remain stable. Consequently, all bifurcations identified in Fig.~\ref{Fig:CPB_Jac} and Table~\ref{Tab:Hopf_Bifurc} are Hopf bifurcations with a two-dimensional center subspace.

Hopf bifurcations are broadly classified as supercritical and subcritical. In a supercritical Hopf bifurcation, a stable limit cycle emerges continuously from the fixed point beyond the bifurcation point, hence the prefix super. In contrast, a subcritical Hopf bifurcation is associated with an unstable limit cycle that exists already before the bifurcation. In this case, the stable fixed point coexists with another attractor, such as a stable limit cycle created at an earlier Hopf bifurcation. The unstable limit cycle forms the separatrix delimiting the basin of attraction of the fixed point. As the bifurcation point is approached, this basin shrinks and eventually vanishes when the unstable limit cycle merges with the fixed point. 

Distinguishing between these two cases requires going beyond linear stability analysis based on the spectrum of $\mathbb{J}(\bm{X}_0)$. Instead, one must analyze the reduced dynamics on the center manifold associated with the fixed point. Center manifold theory guarantees the existence of a locally invariant manifold tangent to the center subspace at the bifurcation point. For a Hopf bifurcation, this manifold is two-dimensional. It may be viewed as the slow sector of the dynamics, containing the periodic orbits that emerge near the fixed point. In contrast, trajectories along the stable and unstable directions approach or depart from the fixed point exponentially fast and therefore do not govern the critical dynamics. 

As discussed in Appendix~\ref{App_PB_Norm_Form}, the flow on the center manifold in the vicinity of the fixed point, after shifting the origin to $\bm{X}_0$, in the polar coordinates $(r,\theta)$ takes the form
\begin{subequations}
\begin{align}
    \dot{r} &= \beta\zeta(\eta)r + a_1 r^3 + \mathcal{O}\left(r^5\right), \label{PB_Radial}\\
    \dot{\theta} &= \omega(\eta) + b_1 r^2 + \mathcal{O}\left(r^4\right), 
\end{align}
\label{PB_Normal_Form}
\end{subequations}
where $\omega(\eta)$ denotes the $\eta$-dependent angular frequency, $a_1$ is the first Lyapunov coefficient in the chosen normal-form convention, $b_1$ characterizes the nonlinear frequency shift. Additionally, $\beta$ and $\zeta(\eta)$ are defined as follows: 
\begin{equation}
    \beta = \frac{d \; \mathrm{Re} (\mu)}{d\eta}\Bigg|_{\eta = \eta_{c}}, \qquad \zeta(\eta) = \left(\eta - \eta_c\right) \equiv \delta\eta_c.
    \label{beta_BK}
\end{equation}
Here $\mu$ is the critical eigenvalue of the Jacobian.  The nature of the Hopf bifurcation is fully determined by the radial part of the Poincar\'{e}-Birkhoff normal form \eqref{PB_Radial}. 

The limit-cycle solutions near the Hopf point are obtained by setting $\dot{r} = 0$ in Eq.~\eqref{PB_Radial}. This gives the trivial solution $r = 0$, corresponding to the fixed point, and the nonzero solution
\begin{equation}
    r = r_{H} = \sqrt{-\frac{\beta\delta\eta_c}{a_{1}}},
    \label{LC_Ampltd}
\end{equation}
which corresponds to a small-amplitude limit cycle near the Hopf bifurcation. The nonzero solution exists only when $r_H^2>0$ -- i.e., when $\beta\delta\eta_c$ and $a_1$ have opposite signs.

For $a_1<0$, the condition for existence is $\beta\delta\eta_c>0$. The limit cycle therefore appears on the side of the bifurcation where the fixed point has lost stability. This corresponds to a supercritical Hopf bifurcation. For example, if $\beta>0$, this occurs for $\delta\eta_c>0$. In contrast, for $a_1>0$, the condition for existence is $\beta\delta\eta_c<0$. The limit cycle then exists on the side where the fixed point is still stable, corresponding to a subcritical Hopf bifurcation with $\beta>0 \implies \delta\eta_c<0$.

Thus, once the radial equation is written in the standard Poincar\'{e}-Birkhoff form, the sign of $a_1$ alone determines whether the Hopf bifurcation is supercritical or subcritical. The sign of $\beta$ specifies how this normal form unfolding is oriented with respect to the physical parameter $\eta$, i.e., whether the Hopf point is approached from below or from above in $\eta$.   

To determine the stability of these limit cycles, we linearize Eq.~\eqref{PB_Radial} about a stationary solution $r_0 = 0$. Writing $\delta r \equiv (r - r_0)$, we obtain   
\begin{multline}
    \frac{d \delta r}{dt} = \left(\beta\delta\eta_c + 3a_1r_0^2\right)\delta r \\    
= \begin{cases}
\beta\delta\eta_c\delta r, & \text{if $r_0 = 0$}, \\
-2\beta\delta\eta_c\delta r, & \text{if $r_0 = r_H$.}
\end{cases}
\label{LC_Lin_Stab}
\end{multline}
Recall that $r_0 = 0$ corresponds to the fixed point, while $r_0 = r_H$ corresponds to the limit cycle. For a supercritical Hopf bifurcation, $(a_1<0)$, the existence condition $r_H\in\mathbb{R}$ implies $\beta\delta\eta_c>0$. Conversely, for a subcritical Hopf bifurcation, the limit cycle exists before the bifurcation, i.e., for $\beta\delta\eta_c<0$. This then means that when the limit cycle exists, i.e., $r_H \in \mathbb{R}$, it is unstable.

\section{Numerical results}\label{numerics} 

\begin{table}[h!]
\centering
\begin{tabular}{| c | c | c |} 
 \hline
 Value of $E_{J} = \Gamma$ & $a_1$ at $\eta_{c1}$ & $a_1$ at $\eta_{c2}$ \\ [0.5ex] 
 \hline\hline
 150.0 & -0.00012 (0.87) & 0.14 (11.99) \\ 
 10.0 & -0.00012 (0.87) & 0.0080 (12.19) \\
 1.0 & -0.00019 (1.06) & -0.0056 (13.29) \\ [1ex] 
 \hline
\end{tabular}
\caption{Summary of the numerical continuation and bifurcation analysis performed using \texttt{BifurcationKit.jl} \cite{Veltz}. As the parameter $\eta$ is increased, the system undergoes two Hopf bifurcations at the critical values $\eta_{c1}$ and $\eta_{c2}$. The first column lists the values of $E_J = \Gamma$, while the second and third columns report the coefficient $a_1$ [cf. Eq.~\eqref{PB_Normal_Form}] evaluated at the corresponding critical values of $\eta$, shown in parentheses. The sign of $a_1$ reveals a supercritical Hopf bifurcation followed by an inverse subcritical Hopf bifurcation in the adiabatic regime $E_J ( = \Gamma) \gg \omega_0$. As the adiabaticity condition is relaxed and $E_J ( = \Gamma)$ becomes comparable to $\omega_0$, the second bifurcation changes character, resulting in a supercritical Hopf bifurcation followed by an inverse supercritical Hopf bifurcation. }
\label{Tab:Hopf_Bifurc}
\end{table}

Our numerical results are summarized in Fig.~\ref{Fig:CPB_Jac} and Table~\ref{Tab:Hopf_Bifurc}. As the coupling parameter $\eta$ is increased from 0.01 to 15.0, the spectrum of the Jacobian $\mathbb{J}(\bm{X}_0; \eta)$ reveals two distinct Hopf bifurcations. We denote the corresponding critical values by $\eta_{c1}$ and $\eta_{c2}$. The stability of the fixed point shows reentrant behavior in that it first loses stability at $\eta_{c1}$ and then regains stability at $\eta_{c2}$. This behavior persists in both the adiabatic and nonadiabatic regimes of the CPB dynamics.  

In Table~\ref{Tab:Hopf_Bifurc}, we provide the numerical results obtained using \texttt{BifurcationKit.jl}. First, we note that the values of $\eta_{c1}$ and $\eta_{c2}$ matched up to two significant digits with the ones obtained from Fig.~\ref{Fig:CPB_Jac}. From the signs of the first Lyapunov coefficients, we conclude that in the adiabatic regime we first have a supercritical, and then a subcritical Hopf bifurcation. Deep in the nonadiabatic regime, where retardation effects become quite significant, we observe two successive supercritical Hopf bifurcations. 

As a consequence, in the adiabatic regime, an infinitesimal limit cycle emerges immediately beyond $\eta_{c1}$. This limit cycle subsequently grows in amplitude as $\eta$ is increased. The fixed point regains stability through an inverse subcritical Hopf bifurcation. As discussed in Appendix~\ref{App_PB_Norm_Form}, the direction of the stability change at a Hopf bifurcation is determined by the sign of $\beta = (d \;\mathrm{Re}\; \mu/d\eta)$ evaluated at the critical value of $\eta$. This, in turn, is fully consistent with the spectral evolution shown in Fig.~\ref{Fig:CPB_Jac}. Owing to the subcritical nature of the second Hopf bifurcation, the fixed point $\bm{X}_0$ coexists with the stable limit cycle generated at the first supercritical Hopf bifurcation for $\eta>\eta_{c2}$ in the immediate vicinity of criticality. 

Finally, the character of the second Hopf bifurcation changes in the nonadiabatic regime. In particular, the inverse supercritical Hopf bifurcation at $\eta_{c2}$ gives rise to an infinitesimal limit cycle for $\eta<\eta_{c2}$ in the immediate vicinity of the critical point. The large-amplitude limit cycle generated at the first supercritical Hopf bifurcation at $\eta = \eta_{c1}$ continues to persist in this region. Consequently, for $\eta\lesssim\eta_{c2}$, the system exhibits coexistence between two distinct limit cycles. Initial conditions chosen sufficiently close to the fixed point $\bm{X}_0$ evolve toward the small-amplitude limit cycle, whereas initial conditions farther away are attracted to the large-amplitude limit cycle.

\section{Discussion}\label{discussion}

The nonlinear electromechanical instability analyzed in this work is controlled by the dimensionless parameter $\eta$, which is physically controlled by the external electric field $\mathcal{E}$, which triggers instability in the considered superconducting NEM device. 
The field is responsible for the pumping of nanovibrations through the interplay between the Andreev current and the position-dependent Josephson coupling, and determines the onset of the nanovibrations at critical point, which occurs at $\eta_{c1}\approx 1$ (see Table.~\ref{Tab:Hopf_Bifurc}). As a result, increasing $\eta$  causes the loss of stability of the fixed point and gives rise to self-sustained vibrations through the Hopf bifurcation at $\eta_{c1}$. In both the adiabatic and nonadiabatic regimes, this bifurcation is found to be supercritical, leading to the continuous emergence of a stable limit cycle from the fixed point. 

Further increase in $\eta$ (and thus in $\mathcal{E}$) produces a shift of the electrostatic energy of the island away from the charge degeneracy. As the charge degeneracy is needed for Coulomb deblocking of a single Cooper-pair tunneling, its removal leads to the reduction of the Cooper-pair tunneling. Consequently, at larger $\eta$, more points of the island trajectory are under conditions of the restored Coulomb blockade of the single Cooper-pair tunneling. This suppresses the Josephson coupling and the corresponding Josephson force acting on the mechanical motion of the island. This competition between the pumping and the suppression of the Cooper-pair tunneling explains the reentrant stability observed in Fig.~\ref{Fig:CPB_Jac}. Once $\eta$ exceeds a second critical value $\eta_{c2}$, the field-induced Coulomb blockade suppresses the pumping and the fixed point regains stability. 

An important result of the present work is that the character of the second Hopf bifurcation depends on the adiabaticity of the island. 
In the adiabatic regime, $E_J, \Gamma\gg \hbar \omega_0$, the second transition is an inverse subcritical Hopf bifurcation. Consequently, the stable limit cycle coexists with the fixed point in the vicinity of $\eta_{c2}$. In contrast, in the nonadiabatic regime, $E_J, \Gamma \sim \hbar \omega_0$, the second transition becomes an inverse supercritical Hopf bifurcation. In this case, for $\eta\lesssim\eta_{c2}$, two limit cycles coexist. 

These findings extend the previous adiabatic analysis of Ref.~\cite{Park26} and demonstrate that self-sustained vibrations persist beyond the adiabatic regime. The existence of the nonadiabatic instability suggests that Cooper-pair transport via a movable island is not restricted to slow mechanical motion, and may provide a robust mechanism for Cooper-pair shuttling over a broad range of parameters. More generally, the present results reveal a rich nonlinear dynamical phase diagram arising from the interplay of Andreev tunneling, Josephson coupling, and mechanical motion, and establish the movable Cooper-pair box as a promising platform for exploring nonlinear superconducting nanoelectromechanics.

\begin{acknowledgments}
We thank Boris L. Altshuler for valuable discussions that inspired the initiation of this work. This work was supported by the National Research Foundation of Korea (NRF) grant funded by the Korea government (MSIT) (Grants No.~RS-2025-16070482, RS-2025-25464760, RS-2026-25519864, RS-2025-25446099, RS-2023-NR119928, RS-2025-03392969). This work was supported by APCTP-2026-S04. This work was supported by the Global Joint Research Program funded by the Pukyong National University(202506520001). L.~Y.~Gorelik and R.~I.~Shekhter acknowledge the hospitality of the PCS at IBS, Republic of Korea, where part of this work was supported by IBS funding No.~IBS-R024-D1. S.~P.~acknowledges the support from the Institute for Basic Science (IBS) in the Republic of Korea through the project IBS-R024-Y4.
\end{acknowledgments}

\appendix
\section{Center Manifold Reduction and the Poincar\'{e}-Birkhoff Normal Form}
\label{App_PB_Norm_Form}

The vector field describing the reduced dynamics on the center manifold is known as the Poincar\'{e}-Birkhoff normal form. Here, we briefly summarize the derivations presented in Refs.~\cite{Crawford,Patra_1} and adapt them to the CPB equations \eqref{DL_EOM}. To this end, we  shift the origin of the coordinate system to $\bm{X}_0$, and rewrite Eq.~\eqref{DL_EOM} as
\begin{equation}
\dot{\bm{X}} = \bm{g}(\bm{X}; \xi),
\label{DL_EOM_VF}
\end{equation}
where $\bm{X} \equiv \left( x - x_0, v_{x} - v_{x0}, y - y_0, v_{y} - v_{y0}, \bm{m} - \bm{m}_0 \right)$ denotes the 7D state vector comprising the mechanical and qubit degrees of freedom, and $\bm{g}(\bm{X}_0; \xi)$ is the corresponding vector field. The parameter $\xi$ serves as the bifurcation parameter. For convenience, we define $\xi = \pm \eta$ such that increasing $\xi$ always corresponds to approaching the Hopf bifurcation, irrespective of the direction in which the critical point is encountered in $\eta$. The fixed point $\bm{X}_0$ is the solution of $\bm{g}(\bm{X}_0; \xi) = 0$.

In this section, we outline the derivation of the Poincar\'{e}-Birkhoff normal form for the standard case in which the fixed point loses stability as the control parameter \(\xi\) is increased. In the main text, however, we also encounter the opposite situation, where the fixed point gains stability upon increasing the control parameter. To treat this case consistently, one must keep track of the crossing direction of the critical eigenvalues by retaining an additional coefficient \(\beta\); see Sec.~\ref{App_PALC}. 

Linearizing Eq.~\eqref{DL_EOM_VF}, we obtain 
\begin{equation}
\frac{d\delta\bm{X}}{dt} = \mathbb{J}(\bm{X}_0; \xi)\cdot \delta\bm{X},
\label{DL_EOM_Jac}
\end{equation}
where $\mathbb{J}(\bm{X}_0; \xi)$ is the Jacobian matrix of the original CPB equations \eqref{DL_EOM}, evaluated at the fixed point $\bm{X}_0$. Here we have used the relation 
\begin{equation}
    \frac{\partial \bm{g(\bm{X})}}{\partial \bm{X}} \Bigg|_{\bm{X} = \bm{0}} = \mathbb{J}(\bm{X}_0; \xi),
\end{equation}
where $\bm{X} = \bm{0}$ corresponds to the fixed point in the shifted coordinate system. At a Hopf bifurcation, a complex-conjugate eigenvalue pair of $\mathbb{J}(\bm{X}_0; \xi)$ becomes purely imaginary, $\pm i\omega(\xi_c)$, while all remaining eigenvalues satisfy $\mathrm{Re} \; \mu < 0$. The eigenvectors associated with this critical eigenvalue pair span the two-dimensional center subspace, which is tangent to the center manifold at the fixed point.

To perform the center-manifold reduction, we first introduce a \textit{real} linear transformation $\bm{X}\mapsto\bm{X}'$ that separates the critical and stable directions \cite{Footnote_1}. In the transformed coordinates, the dynamics takes the form
\begin{multline}
\dot{\bm{X}}'
=
\begin{pmatrix}
\mathbb{A}_c & \mathbb{O} \\
\mathbb{O}^\mathrm{T} & \mathbb{A}_{\perp}
\end{pmatrix}
\bm{X}'
\\ +
\textrm{second- or higher-order terms},
\label{Linear_Block_Form}
\end{multline}
where $\mathbb{A}_c$ is the $2\times 2$ block associated with the critical complex-conjugate eigenvalue pair, $\mathbb{A}_{\perp}$ is the $5 \times 5$ block describing the remaining stable directions, and $\mathbb{O}$ is a $2\times 5$ matrix of zeros. This block structure naturally separates the two coordinates associated with the center subspace from the transverse stable coordinates. We therefore partition the transformed state vector $\bm{X}^\prime$ as 
\begin{equation}
    \bm{X}^\prime \equiv \begin{pmatrix} \bm{X}^\prime_c \\ \bm{X}^\prime_{\perp} \end{pmatrix}, \qquad \bm{X}^\prime_c \equiv (X^\prime_1, X^\prime_2), 
\label{Center_Stable_Decomp}
\end{equation}
where $\bm{X}^\prime_c$ denotes the coordinates associated with the center directions and locally parameterizes the center manifold, whereas $\bm{X}^\prime_{\perp}$ contains the coordinates associated with the stable directions transverse to it.

Near the fixed point, the center manifold is represented as \cite{Crawford, Patra_1}
\begin{equation}
X'_i=h_i(X'_1,X'_2),
\qquad i=3,\ldots,7.
\label{Center_MF}
\end{equation}
Equivalently, a point on the center manifold can be written as $\left(\bm{X}^{\prime}_c, \bm{h}_c\left(\bm{X}^{\prime}_c\right)\right),$ where 
\begin{equation}
    \bm{h}_c\left(\bm{X}^{\prime}_c\right) = \begin{pmatrix} h_3(X'_1,X'_2) \\ h_4(X'_1,X'_2) \\ \vdots \\ h_7(X'_1,X'_2) \end{pmatrix}
    \label{Map_Center_to_Stable}
\end{equation}
maps the center subspace to the stable subspace. Thus, in a sufficiently small neighborhood of the fixed point, the center manifold is locally parameterized by the two center manifold coordinates $\bm{X}^{\prime}_c$, while the remaining coordinates are uniquely determined by Eq.~\eqref{Center_MF}. 

Note that the functional form of $\bm{h}_c\left(\bm{X}^\prime_c\right)$ is constrained by two conditions: (1) the center manifold passes through the fixed point, and (2) it is tangent to the center subspace at that point. Using these conditions (see, e.g., Appendix B of Ref.~\cite{Patra_1}), one immediately deduces that the functions $h_{i}$ have the following form:
\begin{multline}
    h_i(X'_1,X'_2) = h_{i1}\left(X^\prime_{1}\right)^2 + h_{i2}\left(X^\prime_{2}\right)^2 + h_{i3}X^\prime_{1}X^\prime_{2} \\ 
    + \textrm{higher-order terms,} \qquad i = 3, \ldots, 7.
    \label{Map_Center_to_Stable_Fnl_Form}
\end{multline}
These \textit{real} coefficients $h_{ij}$, for $i = 3, \ldots, 7$  and $j=1, 2, 3$ will be obtained using Eq.~\eqref{Stable_Center_MF}.

In the coordinates introduced using Eqs.~\eqref{Linear_Block_Form}, \eqref{Center_Stable_Decomp}, \eqref{Center_MF}, and \eqref{Map_Center_to_Stable}, the CPB dynamics \eqref{DL_EOM_VF} takes the form 
\begin{subequations}
\begin{align}
    \begin{pmatrix}
        \dot{X}^\prime_{1} \\ \dot{X}^\prime_{2}
    \end{pmatrix} &= \underbrace{\begin{pmatrix}
        \zeta(\xi) & \omega(\xi) \\
        -\omega(\xi) & \zeta(\xi)
    \end{pmatrix}}_{\mathbb{A}_c}
    \begin{pmatrix}
        X^\prime_{1} \\ X^\prime_{2}
    \end{pmatrix} \nonumber \\ 
    & \hspace{2cm}+ \begin{pmatrix}
        R_{1}\left(\bm{X}^\prime_c, \bm{h}_c\right) \\ R_{2}\left(\bm{X}^\prime_c, \bm{h}_c\right)
    \end{pmatrix}, \label{DL_EOM_CMF_Proj} \\
    \dot{\bm{X}}^\prime_{\perp} &= \underbrace{\begin{pmatrix} 
\zeta^\prime(\xi) & \omega^\prime(\xi) & 0 & 0 & 0 \\ 
-\omega^\prime(\xi) & \zeta^\prime(\xi) & 0 & 0 & 0 \\ 
0 & 0 & \zeta^{\prime\prime}(\xi) & \omega^{\prime\prime}(\xi) & 0 \\
0 & 0 & -\omega^{\prime\prime}(\xi) & \zeta^{\prime\prime}(\xi) & 0 \\
0 & 0 & 0 & 0 & \mu_{7} 
\end{pmatrix}}_{\mathbb{A}_{\perp}} 
\begin{pmatrix}
    h_{3} \\ h_{4} \\ h_{5} \\ h_{6} \\ h_{7}
\end{pmatrix} \nonumber \\
& \hspace{2cm} + 
\begin{pmatrix}
    R_{3}\left(\bm{X}^\prime_c, \bm{h}_c\right) \\ R_{4}\left(\bm{X}^\prime_c, \bm{h}_c\right) \\ 
    R_{5}\left(\bm{X}^\prime_c, \bm{h}_c\right) \\ 
    R_{6}\left(\bm{X}^\prime_c, \bm{h}_c\right) \\ R_{7}\left(\bm{X}^\prime_c, \bm{h}_c\right)
\end{pmatrix},
\label{DL_EOM_SMF_Proj}
\end{align}
\label{DL_EOM_CMF}
\end{subequations}
where $\zeta \pm i\omega$ with $\zeta(\xi_c) = 0$ are the critical eigenvalues associated with the center manifold, $\zeta^\prime \pm i\omega^\prime, \zeta^{\prime\prime} \pm i\omega^{\prime\prime}$, and $\mu_7$ correspond to the stable directions, and $R_i$ denote the nonlinear contributions to the transformed vector field. Next, substituting Eq.~\eqref{Center_MF} into the equations for the stable coordinates -- i.e., in Eq.~\eqref{DL_EOM_SMF_Proj} -- yields the center-manifold invariance condition,
\begin{multline}
\begin{pmatrix}
    \frac{\partial \bm{h}_c}{\partial X^\prime_{1}} & \frac{\partial \bm{h}_c}{\partial X^\prime_{2}} 
\end{pmatrix}
\begin{bmatrix}
    \mathbb{A}_c
    \begin{pmatrix}
        X^\prime_{1} \\ X^\prime_{2}
    \end{pmatrix} + \begin{pmatrix}
        R_{1} \\ R_{2}
    \end{pmatrix}
\end{bmatrix} \\
= \mathbb{A}_{\perp} 
\begin{pmatrix}
    h_{3} \\ \vdots \\ h_{7}
\end{pmatrix} + 
\begin{pmatrix}
    R_{3} \\ \vdots \\ R_{7}
\end{pmatrix},
\label{Stable_Center_MF}
\end{multline}
which determines the quadratic coefficients of $\bm{h}_c$, namely those multiplying $\left(X^\prime_{1}\right)^2,\left(X^\prime_{2}\right)^2$ and $X^\prime_{1}X^\prime_{2}$. These coefficients are sufficient to construct the Poincar\'{e}-Birkhoff normal form to cubic order and thereby determine the nature of the Hopf bifurcation. 

Substituting the resulting approximation for $\bm{h}_c\left(\bm{X}^\prime_c\right)$ into Eq.~\eqref{DL_EOM_CMF_Proj}, we obtain the flow on the center manifold. The next step is to eliminate nonessential nonlinearities through a sequence of near-identity transformations, 
\begin{equation}
\bm{X}^\prime \mapsto\bm{\mathcal{X}} = \bm{X}^\prime + \phi^{(k)}(\bm{X}^\prime),
\label{NIT}
\end{equation}
where $\phi^{(k)}$ is a homogeneous polynomial map of degree $k$ with $\phi^{(k)}(a\bm{X}^\prime) = a^k \phi^{(k)}(\bm{X}^\prime)$, where $a$ is a real number. Such transformations remove all nonresonant nonlinear terms at order $k$ without affecting the essential nonlinearities at lower orders. Repeating this procedure order by order yields the Poincar\'{e}-Birkhoff normal form
\begin{multline}
    \begin{pmatrix}
        \dot{z} \\ \dot{\bar{z}}
    \end{pmatrix} = 
    \begin{pmatrix}
        \zeta(\xi) + i\omega(\xi) & 0 \\
        0 & \zeta(\xi)-i\omega(\xi)
    \end{pmatrix} \begin{pmatrix}
        z \\ \bar{z}
    \end{pmatrix} \\
    + \sum_{k=1}^{\infty} \begin{pmatrix}
        \alpha_k z |z|^{2k}\\ \bar{\alpha}_k \bar{z} |z|^{2k}
    \end{pmatrix},
    \label{PB_NF_1}
\end{multline}
where $z$ and $\bar{z}$ denote the complex coordinates on the center manifold, and $\alpha_k$ and $\bar{\alpha}_k$ are the coefficients of $\begin{pmatrix} z |z|^{2k} & 0\end{pmatrix}^\mathrm{T}$ and $\begin{pmatrix} 0 &  \bar{z}|z|^{2k}\end{pmatrix}^\mathrm{T}$ respectively. Expressing this coordinate in its polar form, $z = re^{i\theta}$, transforms Eq.~\eqref{PB_NF_1} into the standard amplitude-phase representation of the Poincar\'{e}-Birkhoff normal form: 
\begin{subequations}
\begin{align}
    \dot{r} &= r\left[ \zeta(\xi) + \sum_{k = 1}^{\infty} a_k r^{2k} \right], \\
    \dot{\theta} &= \omega(\xi) + \sum_{k = 1}^{\infty} b_k r^{2k},
\end{align}
\label{PB_Normal_Form_Append}
\end{subequations}
where 
\begin{equation}
    a_k = \mathrm{Re}\left(\alpha_k\right), \qquad b_k = - \mathrm{Im}\left(\alpha_k\right).
    \label{ak_BK}
\end{equation}
The type of the Hopf bifurcation is then determined by the first Lyapunov coefficient $a_1$.

\subsection{Continuation of Equilibria Using BifurcationKit.jl}
\label{App_PALC}

Additionally, we must clarify that the numerical bifurcation analysis is carried out using \texttt{BifurcationKit.jl}, which follows the 8D equilibrium branch by \texttt{Pseudo-Arclength Continuation} (\texttt{PALC}). The continued equilibrium branch is parameterized as $(\bm{X}_0(s), \eta(s))$, where $s$ is an auxiliary continuation parameter corresponding to the pseudo-arclength along the branch. This avoids the ambiguity of choice in the continuation parameter $\xi = \pm \eta$. Accordingly, the center-manifold dynamics near criticality can be written as
\begin{subequations}
\begin{align}
    \dot{r} &= r\left[  \beta\zeta(\eta) + \sum_{k = 1}^{\infty} a_k r^{2k} \right], \\
    \dot{\theta} &= \omega(\eta) + \sum_{k = 1}^{\infty} b_k r^{2k},
\end{align}
\label{PB_Normal_Form_BK}
\end{subequations}
where 
\begin{equation}
    \beta = \frac{d \; \mathrm{Re} (\mu)}{d\eta}\Bigg|_{\eta = \eta_{H}}, \qquad \zeta(\eta) = \left(\eta - \eta_c\right).
    \label{beta_BK_Append}
\end{equation}
Here the dot denotes differentiation with respect to the physical time $t$.


\begin{thebibliography}{99}
\bibitem{Footnote_1} An example of such linear transformation is given in Eq.~(B6) of ref.~\cite{Patra_1}.
\end{thebibliography}
\end{document}